\documentclass[%
 reprint,
 amsmath,amssymb,
 aps,
]{revtex4-1}

\usepackage{graphicx}% Include figure files
\usepackage{dcolumn}% Align table columns on decimal point
\usepackage{bm}% bold math
\begin{document}

\preprint{APS/123-QED}

\title{Deformed black hole immersed in dark matter spike}% Force line breaks with \\

\author{Zhaoyi Xu}
 \email{xuzy@ihep.ac.cn}
 \affiliation{%
Key Laboratory of Particle Astrophysics,
Institute of High Energy Physics, Chinese Academy of Sciences,
Beijing 100049, China.\\
College of Physics, Guizhou University, Guiyang 550025, China. 
}%

\author{Jiancheng Wang}%
\affiliation{%
Yunnan Observatories, Chinese Academy of Sciences, 396 Yangfangwang, Guandu District, Kunming, 650216, China.\\
 University of Chinese Academy of Sciences, Beijing 100049, China.\\
 Key Laboratory for the Structure and Evolution of Celestial Objects, Chinese Academy of Sciences, 396 Yangfangwang, Guandu District, Kunming 650216, China.
}%

%\affiliation{
%\affiliation{
%\affiliation{%

\date{\today}% It is always \today, today,

\begin{abstract}
If a lot of dark matter particles accumulate near the black hole, then the chances of detecting dark matter signals near a black hole are greatly increased. These effects may be observed by the Event Horizon Telescope (EHT), Tianqin project, Taiji project, Laser Interferometer Space Antenna (LISA) and Laser Interferometer Gravitational-Wave Observatory (LIGO). In this work, we explore the effects of dark matter spikes on black hole space-time. For the Schwarzschild-like black hole case, we consider Newton$'$s approximation and perturbation approximation. This makes it possible to use Xu$'$s method to solve the Einstein field equation, and extend Schwarzschild-like black hole to Kerr-like black hole (BH) via Newman-Janis (NJ) algorithm. By analyzing the dark matter spike on the black hole event horizon (EH), stationary limit surfaces (SLS), ergosphere and energy-momentum tensors (EMT), we found that compared with the dark matter halo, the dark matter spike would have a higher effect on the black hole by several orders of magnitude. Therefore, if there is a dark matter spike near the black hole, it is very possible to test the dark matter model through gravitational wave (GW) observation and EHT observation.

\end{abstract}

%\keywords{Dark matter spike, Deforemed black hole, Newman-Janis algorithm}%Use showkeys class option if keyword
\maketitle

%\tableofcontents

\section{Introduction}
So far, there have been many observational evidences for the existence of dark matter (DM)\citep{2017ARA&A..55..343B,2016A&A...594A..13P,2016MNRAS.459.3040G}, such as the cosmic microwave background radiation (CMB), baryon acoustic oscillations (BAO), 
spiral galaxy rotation curve (RC), mass-luminosity ratio of elliptical galaxy (MLR-EG), etc. Based on these observational facts, astronomers have proposed a series of DM models. One of the most popular is the cold dark matter (CDM) model \citep{1996ApJ...462..563N,1997ApJ...490..493N}. 
Although CDM has achieved great success in the large-scale structure of the universe and the formation of galaxies, it has encountered many contradictions in the small-scale observation \citep{2017ARA&A..55..343B}.
These questions cover many aspects of the small scale of galaxies.
For instance, the core/cusp (CC) problem \citep{2001astro.ph..9392K,2015MNRAS.452.3650O}, the Too Big to Fail (TBTF) problem \citep{2011MNRAS.415L..40B,2012MNRAS.422.1203B} and the missing satellite problem (MSP) \citep{2015ApJ...813..109D}. Recently, with the Event Horizon Telescope (EHT)$'$s observations of black hole (BH) shadows and Laser Interferometer Gravitation Wave Observatory (LIGO)$'$s observations of gravitational waves (GW), the existence of BHs in the universe is almost universally accepted \citep{2019ApJ...875L...1E,2016PhRvL.116f1102A}. Therefore, many physicists have begun to study the interaction between BHs and DM \citep{1999PhRvL..83.1719G,2013PhRvD..88f3522S,2017PhRvD..95f4015X,2018JCAP...09..038X,2020PhRvD.101b4029X}. The research on these aspects is mainly divided into two aspects: the DM distribution near BHs and the change of BH metric by DM. 

Supermassive black holes (SMBH) have been observed at the centers of galaxies, and intermediate-mass ratio inspiral (IMRI) systems have also been observed in globular clusters (GC) \citep{2019A&A...625L..10G,2011NewAR..55..166F,2009Natur.460...73F}. For such cases, due to the existence of the BH, the strong gravitational potential of the BH concentrates a large amount of DM particles near the BH horizon \citep{1999PhRvL..83.1719G}. Calculations show that these DM particles form spike distribution. In general, the DM density increases by orders of magnitude due to the BH$'$s gravitational field. Therefore, if DM particles can annihilate into gamma-ray radiation, the intensity of gamma-ray radiation near the BH will increase greatly, which provides a good opportunity for us to detect the DM annihilation signal. In 1999, Gondolo-Silk (GS) first theorized the spike  structure of DM near a BH \citep{1999PhRvL..83.1719G}. In this model, they use adiabatic and Newtonian approximations, and they found that the DM spike is polynomial form. Based on this result, they proposed to use a power-law  distribution to describe the spike distribution, which has been widely used to study various processes involving the distribution of DM near BHs. The GS distribution is valid for the space-time region far from the BH event horizon. But if you take into account the density of DM near the BH, the result is different. Sadeghian-Ferrer-Will(SFW) have studied the spike structure for full relativity approximation in 2013 \cite{2013PhRvD..88f3522S}. They found that when the Newtonian approximation is removed, the density distribution of DM around the Schwarzschild BH changes significantly. The DM density disappeared in 2$R_{S}$, instead of 4$R_{S}$. The peak density of DM is about 15$\%$ higher than Newtonian approximation. These results may have important implications for observations. 

If there are DM particles in the vicinity of the BH, then those DM particles could change the BH$'$s metric. In 2003, Kiselev obtained a BH solution with a specific matter condition \citep{2003CQGra..20.1187K}. Depending on the equation of state, this exact solution can describe the change of the BH metric caused by DM, dark energy (DE) and cosmological constant \citep{2018JCAP...10..046X}. These spherical symmetry BH solution have been generalized to the Kerr-like BH and Kerr-Newman (KN)-like BH case \citep{2015arXiv151201498T,2017PhRvD..95f4015X}. In the case of violating energy-momentum tensor conversation (EMC), the Kiselev-like BH solution has been studied, and they found that the EMC of strong gravity can have a significant effect on the BH metric \citep{2018EPJC...78..513X}. For the Newtonian approximation, disformed BH metric in DM halo has been discussed \citep{2018JCAP...09..038X,2020PhRvD.101b4029X}. Recently, Event Horizon Telescope (EHT) observed the BH shadow of the SMBH in the center of M87, it observed the approximate size and outline of the BH shadow \citep{2019ApJ...875L...1E}. With the accumulation of EHT observation data, it is possible to use BH shadow to study the matter distribution and characteristics near the BH.

Compared to the observational accuracy of EHT, the effect of DM halo on the BH metric (and BH shadow) is still small. But if it was a DM spike, the results would be different. Therefore, it is meaningful to study the effect of DM spike on BH metric. Here, we will discuss these issues. We derive the deformed BH metric in the DM spike background, and analyze the properties of BH. 
The structure of this paper is as follows: In Sec. \ref{dms}, we introduce the DM spike near BH and in Sec. \ref{metrics} we derive the deformed BH metric in the DM spike background. We obtain the deformed Kerr-like BH immersed DM spike via Newman-Janis (NJ) algorithm in Sec. \ref{metric-kerr}. Sec. \ref{sum-} is the summary. The G=c=1.

\section{Dark matter spike near black hole}
\label{dms}
In the CDM model, the DM density of galactic is given by Navarro-Frenk-White (NFW) profile \citep{1996ApJ...462..563N,1997ApJ...490..493N}. For the self-interaction dark matter (SIDM) model, the DM density is $\rho_{core}=m_{\chi}/(\sigma\upsilon t_{\rm bh})$, where $\sigma\upsilon$ is the scattering cross-section, $m_{\chi}$ is the DM particle mass, $t_{\rm bh}$ is the time which equal to $10^{10}$year \citep{2005PhR...405..279B}. For the fuzzy dark matter (FDM) model, the DM density distribution is Tomas-Feimi (TF) profile (Uniform Density) \citep{C2007Can}. If radial distance $r\rightarrow 0$, these DM profile can be approximated in power law form \citep{1999PhRvL..83.1719G}
\begin{equation}
\rho_{\rm DM}(r)=\rho_{\rm 0}\big(\dfrac{r}{R_{\rm 0}}\big)^{-\alpha_{ini}},
\label{DM-S0}
\end{equation}
where $\rho_{\rm 0}$ is the center density, $R_{\rm 0}$ is the scale radius and $\alpha_{ini}$ is the power-law index. For the NFW profile, the $\alpha_{ini}=1$; for SIDM and TF model, the $\alpha_{ini}=0$. Furthermore, if a SMBH at the center of a galaxy or globular clusters (GC), the DM density have a enhancement effect due to the strong gravitational field of the BH. When the BH-DM system satisfy the adiabatic condition and Newtonian approximations \citep{1999PhRvL..83.1719G}, the DM density become 
\begin{equation}
\rho_{\rm GS}(r)=\rho_{\rm R}\big(1-\dfrac{4R_{S}}{r}\big)^{3}\big(\dfrac{r}{R_{\rm sp}}\big)^{-\alpha},
\label{DM-S1}
\end{equation}
where $\rho_{\rm R}$ is the normalization of the DM density, $R_{S}$ is the Schwarzschild radius of the BH, $R_{\rm sp}$ is the radius of the DM spike, $M_{\rm BH}$ is the initial BH mass and $\alpha$ is the power-law index. The power-law index $\alpha$ of the DM spike, which depends on the initial DM profile ($\alpha=(9-2\alpha_{ini})/(4-2\alpha_{ini})$). If the initial DM profile is NFW, the adiabatic growth of the BH lead to the parameter $\alpha=7/3$ \citep{1997ApJ...490..493N}. If the initial DM profile has a Uniform Density profile (TF and SIDM, etc.), the power-law index become $\alpha=3/2$ \citep{2001PhRvD..64d3504U,1995ApJ...440..554Q}. 

SFW abandoned the Newtonian condition to study the DM spike in the full relativistic situation \citep{2013PhRvD..88f3522S}, they found that the integral form of the DM spike profile is
\begin{widetext}
\begin{equation}
\rho_{\rm FR}(r)=\dfrac{4\pi}{r^{2}}\int^{0}_{-GM_{\rm BH}/r}dE\int^{L_{\rm max}}_{0}LdL\dfrac{f^{'}(E^{'}(E,L),L)}{\sqrt{2E+\dfrac{2GM_{\rm BH}}{r}-\dfrac{L^{2}}{r^{2}}}} ,
\label{DM-S21}
\end{equation}
\end{widetext}
where $L$ is the angular momentum of the DM particles, $E$ is the energy of the DM particles. $L_{\rm max}$ is the maximum angular momentum of the DM particles. $f^{'}(E^{'}(E,L),L)$ is the distribution function. Through numerical calculation, SFW found the following results: (1) the density of DM at zero is $2R_{S}$ instead of $4R_{S}$ in the Newtonian approximation; (2) the peak density of the DM spikes increased by $15\%$. From Fig.1 in the original SFW article, we found that the full relativistic does not change $R_{sp}$. Therefore, TXW suggests describing the DM spike profile as follows \citep{2021ChPhC..45a5110T}
\begin{equation}
\rho_{\rm FR}(r)=\rho_{\rm R}\big(1-\dfrac{2 R_{S}}{r}\big)^{3}\big(\dfrac{r}{R_{\rm sp}}\big)^{-\alpha}.
\label{DM-S2}
\end{equation}
Combining Eq. \ref{DM-S1} and Eq. \ref{DM-S2}, it can be found that from Newtonian approximation to full relativistic case, the DM enhancement effect is continuously strengthened. From TXW work \citep{2021ChPhC..45a5110T}, the DM spike profile around Schwarzschild BH is
\begin{equation}
\rho(r)=\rho_{\rm R}\big(1-\dfrac{kR_{S}}{r}\big)^{3}\big(\dfrac{r}{R_{\rm sp}}\big)^{-\alpha},
\label{DM-S3}
\end{equation}
where $k$ is the DM zero point parameter. For the Newtonian approximation, $k=4$, and full relativistic case $k=2$. For M87 BH-DM spike system, DM spike parameters $\rho_{\rm R}=9.12kpc$ and $R_{\rm sp}=6.9\times 10^{-3}M_{\bullet}/pc^{3}$. For GC BH-DM minspike system, DM spike parameters $\rho_{\rm R}=0.54pc$ and $R_{\rm sp}=6.9\times 10^{-3}M_{\bullet}/pc^{3}$. In natural units, $R_{\rm sp}=1.44\times 10^{7}$, $\rho_{\rm R}=1.12\times 10^{-21}$ for M87 (Fig.\ref{EH100} and \ref{EH200}). $R_{\rm sp}=1.44\times 10^{7}$, $\rho_{\rm R}=1.12\times 10^{-21}$ for GC.

\section{Deformed schwarzschild black hole in dark matter spike}
\label{metrics}
In this section, we derive the deformed schwarzschild black hole metric in the DM spike via the Xu$'$s method \citep{2018JCAP...09..038X}. This method typically consists of two steps: Firstly, in the case of general relativity, the DM space-time metric is constructed based on the DM spike profile; Secondly, by analyzing Einstein$'$s field equation, the approximate solution of BH under DM spike is constructed. It is important to point out that these deformed BH metrics satisfy Newtonian approximate, therefore the metric coefficient $f(r)=g(r)$. On the other hand, in the vicinity of SMBHs, the number of stars should be very small due to the activity of the central BH or the strong gravitational field of the BH, therefore, we only consider the effect of DM spike on BH space-time metric.

For the first step to deformed schwarzschild BH is to known that the mass distribution of DM spike. 
In the case of spherical symmetry space-time, the mass distribution can be given by DM spike density profile (\ref{DM-S3}), and the mass profile is  
\begin{widetext}
\begin{equation}
M_{\rm DM}=4\pi\int^{r}_{kR_{S}}\rho(r')r'^{2}dr' 
=\dfrac{4\pi R^{\alpha}_{\rm sp}\rho_{\rm R}k^{3}R_{S}^{3}(r^{-\alpha}-(kR_{S})^{-\alpha})}{\alpha}-\dfrac{12\pi R^{\alpha}_{\rm sp}\rho_{\rm R}k^{2}R_{S}^{2}(r^{1-\alpha}-(kR_{S})^{1-\alpha})}{\alpha-1}+$$$$
\dfrac{12\pi R^{\alpha}_{\rm sp}\rho_{\rm R}kR_{S}(r^{2-\alpha}-(kR_{S})^{2-\alpha})}{\alpha-2}-\dfrac{4\pi R^{\alpha}_{\rm sp}\rho_{\rm R}(r^{3-\alpha}-(kR_{S})^{3-\alpha})}{\alpha-3}.
\label{DM-S4}
\end{equation}
\end{widetext}
For a test particle located on the equatorial plane in spherical symmetry space-time. According to Newtonian theory, the tangential velocity is determined by the mass distribution of the DM spikes, its tangential velocity $V$ is
\begin{widetext}
\begin{equation}
\begin{aligned}
&V=\sqrt{\dfrac{M_{\rm DM}}{r}}=\Big(\dfrac{4\pi R^{\alpha}_{\rm sp}\rho_{\rm R}k^{3}R_{S}^{3}(r^{-\alpha-1}-(kR_{S})^{-\alpha}/r)}{\alpha}-\dfrac{12\pi R^{\alpha}_{\rm sp}\rho_{\rm R}k^{2}R_{S}^{2}(r^{-\alpha}-(kR_{S})^{1-\alpha}/r)}{\alpha-1}\\
&+\dfrac{12\pi R^{\alpha}_{\rm sp}\rho_{\rm R}kR_{S}(r^{1-\alpha}-(kR_{S})^{2-\alpha}/r)}{\alpha-2}-\dfrac{4\pi R^{\alpha}_{\rm sp}\rho_{\rm R}(r^{2-\alpha}-(kR_{S})^{3-\alpha}/r)}{\alpha-3}\Big)^{\dfrac{1}{2}}.
\label{DM-S5}
\end{aligned}
\end{equation}
\end{widetext}

Generally speaking, the space-time metric is determined by a pure DM spike profile, and the spherical symmetry space-time metric is given by 
\begin{equation}
ds^{2}=-f(r)dt^{2}+\dfrac{dr^{2}}{g(r)}+r^{2}(d\theta^{2}+\sin^{2}\theta d\phi^{2}), 
\label{DM-S6}
\end{equation}
where $f(r)$ and $g(r)$ are the redshift functions and shape functions, respectively. On the equatorial plane of spherical symmetry space-time. The tangential velocity of the test particle is closely related to the space-time metric coefficients, which can be obtained by analyzing the geodesic equation in the case of general relativity (GR). 
The relation between the tangential velocity $V$ of the test particle and the redshift function $f(r)$ is
\begin{equation}
V^{2}=\dfrac{r}{\sqrt{f(r)}}\dfrac{d\sqrt{f(r)}}{dr}=r\dfrac{dln\sqrt{f(r)}}{dr}.
\label{DM-S7}
\end{equation} 
It is worth noting that the restriction condition $f(r)=g(r)$ is used, the corresponding reasons and details are shown in Reference \cite{2018JCAP...09..038X}. Combining equations (\ref{DM-S5}) and (\ref{DM-S7}), we obtain an analytical expression for the space-time metric coefficients $f(r)$ is 
\begin{widetext}
\begin{equation}
\begin{aligned}
&f(r)=g(r)=r^{\dfrac{48\pi R^{\alpha}_{\rm sp}\rho_{\rm R}(kR_{S})^{3-\alpha}}{\alpha(\alpha-1)(\alpha-2)(\alpha-3)}}\times {\rm exp}\Big(-\dfrac{8\pi R^{\alpha}_{\rm sp}\rho_{\rm R}(kR_{S})^{3}}{\alpha^{2}}r^{-\alpha}+\\
&\dfrac{24\pi R^{\alpha}_{\rm sp}\rho_{\rm R}(kR_{S})^{2}}{(\alpha-1)^{2}}r^{1-\alpha}-
\dfrac{24\pi R^{\alpha}_{\rm sp}\rho_{\rm R}kR_{S}}{(\alpha-2)^{2}}r^{2-\alpha}+\dfrac{8\pi R^{\alpha}_{\rm sp}\rho_{\rm R}}{(\alpha-3)^{2}}r^{3-\alpha}\Big).
\label{DM-S8}
\end{aligned}
\end{equation}
\end{widetext}
When DM spike disappears, that is to say $\rho_{\rm R}=0$, the metric coefficients $f(r)$ and $g(r)$ reduce to
\begin{equation}
{\lim_{\rho_{\rm R} \to 0}}f(r)={\lim_{\rho_{\rm R} \to 0}}g(r)=1,
\label{DM-S9}
\end{equation}
this is the flat space-time metric, the usual Minkowski space-time. Results \ref{DM-S9} may also suggest that the space-time metric of DM spike may be a perturbation of Minkowski space-time, which is not very far away from Minkowski space-time. 

For the second step to deformed schwarzschild BH is to known that the solve the Einstein$'$s field equation. According to the Xu$'$s method \citep{2018JCAP...09..038X}. We need to solve the Einstein$'$s field equation in the case of a DM spike distribution and a point mass distribution. If the energy-momentum tensor $T_{\mu\nu}$ is completely determined by the DM spike profile. the the Einstein$'$s field equation is given by 
\begin{equation}
R_{\mu\nu}-\dfrac{1}{2}g_{\mu\nu}R=\kappa^{2}T_{\mu\nu}(\rm DM-spike),
\label{DM-S10}
\end{equation}
where $\kappa^{2}=8\pi$, $R_{\mu\nu}$ is the Ricci curvature tensor, $R$ is the Ricci scalar. Generally, the energy-momentum tensor can be written as $T^{\nu}_{~\mu}=g^{\nu\sigma}T_{\mu\sigma}=diag[-\rho,p_{r},p,p]$. Therefore, the Einstein$'$s field equation become 
%\begin{equation}
%\kappa^{2}T^{t}_{~t}({\rm DM-spike})=g(r)(\dfrac{1}{r}\dfrac{g^{'}(r)}{g(r)}+\dfrac{1}{r^{2}})-\dfrac{1}{r^{2}},\\\\
%\kappa^{2}T^{r}_{~r}({\rm DM-spike})=g(r)(\dfrac{1}{r^{2}}+\dfrac{1}{r}\dfrac{f^{'}(r)}{f(r)})-\dfrac{1}{r^{2}},\\\\
%\kappa^{2}T^{\theta}_{~\theta}({\rm DM-spike})=\kappa^{2}T^{\phi}_{~\phi}({\rm DM-spike})=\dfrac{1}{2}g(r)(\dfrac{f^{''}(r)f(r)-f^{'2}(r)}{f^{2}(r)}+\dfrac{1}{2}\dfrac{f^{'2}(r)}{f^{2}(r)}\\\\
%+\dfrac{1}{r}(\dfrac{f^{'}(r)}{f(r)}+\dfrac{g^{'}(r)}{g(r)})+\dfrac{f^{'}(r)g^{'}(r)}{2f(r)g(r)}).
%\label{DM-S11}
%\end{equation}
\begin{widetext}
\begin{equation}
\begin{aligned}
&\kappa^{2}T^{t}_{~t}({\rm DM-spike})=g(r)(\dfrac{1}{r}\dfrac{g^{'}(r)}{g(r)}+\dfrac{1}{r^{2}})-\dfrac{1}{r^{2}}, \\
&\kappa^{2}T^{r}_{~r}({\rm DM-spike})=g(r)(\dfrac{1}{r^{2}}+\dfrac{1}{r}\dfrac{f^{'}(r)}{f(r)})-\dfrac{1}{r^{2}},\\
&\kappa^{2}T^{\theta}_{~\theta}({\rm DM-spike})=\kappa^{2}T^{\phi}_{~\phi}({\rm DM-spike})=\dfrac{1}{2}g(r)(\dfrac{f^{''}(r)f(r)-f^{'2}(r)}{f^{2}(r)}\\
&+\dfrac{1}{2}\dfrac{f^{'2}(r)}{f^{2}(r)}+\dfrac{1}{r}(\dfrac{f^{'}(r)}{f(r)}+\dfrac{g^{'}(r)}{g(r)})+\dfrac{f^{'}(r)g^{'}(r)}{2f(r)g(r)}).
\label{DM-S11}
 \end{aligned}
\end{equation}
\end{widetext}
According to the algorithm in \cite{2018JCAP...09..038X}, if consider BHs in the Einstein$'$s field equation, the total energy-momentum tensor $T^{\nu}_{~\mu}$ can be written as $T^{\nu}_{~\mu}=T^{\nu}_{~\mu}({\rm BH})+T^{\nu}_{~\mu}({\rm DM-spike})$. From general relativity (GR) and BH physics, the external solution of the Schwarzschild BH corresponds to the vacuum case, and satisfies the condition $T^{\nu}_{~\mu}({\rm BH})=0$. When we consider the Schwarzschild BH and DM spike profile, the total space-time metric may be written as 
\begin{widetext}
\begin{equation}
ds^{2}=-(f(r)+F_{1}(r))dt^{2}+(g(r)+F_{2}(r))^{-1}dr^{2}+r^{2}(d\theta^{2}+\sin^{2}\theta d\phi^{2}),
\label{DM-S12}
\end{equation}
\end{widetext}
where $f(r)$ and $g(r)$ are pure DM metric coefficients (\ref{DM-S8}), the functions $F_{1}(r)$ and $F_{2}(r)$ are determined by BH parameters and DM spike parameters. There may also be a coupling between DM spike and BHs. Based on equation (\ref{DM-S12}), the Einstein$'$s field equation (\ref{DM-S10}) becomes 
\begin{equation}
R^{\nu}_{~\mu}-\dfrac{1}{2}\delta^{\nu}_{~\mu}R=\kappa^{2}T^{\nu}_{~\mu}=\kappa^{2}\big(T^{\nu}_{~\mu}({\rm BH})+T^{\nu}_{~\mu}({\rm DM-spike})\big).
\label{DM-S13}
\end{equation}
Combine the equations (\ref{DM-S11}), (\ref{DM-S12}), (\ref{DM-S13}) and condition $T^{\nu}_{~\mu}({\rm BH})=0$. The field equation (\ref{DM-S13}) can be simplified as
\begin{equation}
(g(r)+F_{2}(r))(\dfrac{1}{r^{2}}+\dfrac{1}{r}\dfrac{g^{'}(r)+F^{'}_{2}(r)}{g(r)+F_{2}(r)})=g(r)(\dfrac{1}{r^{2}}+\dfrac{1}{r}\dfrac{g^{'(r)}}{g(r)}),$$$$
(g(r)+F_{2}(r))(\dfrac{1}{r^{2}}+\dfrac{1}{r}\dfrac{f^{'}(r)+F^{'}_{1}(r)}{f(r)+F_{1}(r)})=g(r)(\dfrac{1}{r^{2}}+\dfrac{1}{r}\dfrac{f^{'(r)}}{f(r)}).
\label{DM-S14}
\end{equation}
Taking Schwarzschild BH as the boundary condition, the general solution is obtained as
\begin{equation}
F_{1}(r)=exp\Big(\int\dfrac{g(r)}{g(r)-\dfrac{2M_{\rm BH}}{r}}(\dfrac{1}{r}+\dfrac{f^{'}(r)}{f(r)})-\dfrac{1}{r} dr\Big)-f(r),$$$$
F_{2}(r)=-\dfrac{2M_{\rm BH}}{r}.
\label{DM-S15}
\end{equation}
Based on the assumption of $f(r)=g(r)$, we find that $F_{1}(r)=F_{2}(r)=-2M_{\rm BH}/r$. Therefore, the disformed  Schwarzschild BH in the DM spike can be written as
\begin{equation}
ds^{2}=-F(r)dt^{2}+G(r)^{-1}dr^{2}+r^{2}(d\theta^{2}+\sin^{2}\theta d\phi^{2}),
\label{DM-S16}
\end{equation}
where functions $F(r)=(r)+F_{1}(r)$ and $G(r)=g(r)+F_{2}(r)$. The final deformed BH metric coefficients are
\begin{widetext}
\begin{equation}
\begin{aligned}
&F(r)=G(r)=r^{\dfrac{48\pi R^{\alpha}_{\rm sp}\rho_{\rm R}(kR_{S})^{3-\alpha}}{\alpha(\alpha-1)(\alpha-2)(\alpha-3)}}\times {\rm exp}\Big(-\dfrac{8\pi R^{\alpha}_{\rm sp}\rho_{\rm R}(kR_{S})^{3}}{\alpha^{2}}r^{-\alpha}+\\
&\dfrac{24\pi R^{\alpha}_{\rm sp}\rho_{\rm R}(kR_{S})^{2}}{(\alpha-1)^{2}}r^{1-\alpha}-\dfrac{24\pi R^{\alpha}_{\rm sp}\rho_{\rm R}kR_{S}}{(\alpha-2)^{2}}r^{2-\alpha}+\dfrac{8\pi R^{\alpha}_{\rm sp}\rho_{\rm R}}{(\alpha-3)^{2}}r^{3-\alpha}\Big)-\dfrac{2M_{\rm BH}}{r}.
\label{DM-S17}
\end{aligned}
\end{equation}
\end{widetext}
It is easy to know that when the DM near the BH disappears $\rho_{\rm R}=0$, the deformed metric (\ref{DM-S17}) degenerates to the Schwarzschild BH metric $F(r)=G(r)=1-2M_{\rm BH}/r$. In the CDM model, the initial DM distribution is NFW profile, and the power index of DM spike is $7/3$. In the SIDM and FDM model (Uniform Density), the DM profile of in the center region is a DM core, and the power index of DM spike is $3/2$. The disformed BH metric under different DM spike can be obtained by simply putting the power exponent into the metric coefficient (\ref{DM-S17}).

\section{Disformed Kerr BH in DM spike}
\label{metric-kerr}
Consideration of BH spins in the context of DM spikes is important for at least a few reasons: (1) consider the BH spin as an important extension of deformed Schwarzschild BH \citep{2017PhRvD..96h3014F}; (2) from the expression of the DM spike, we know that the DM spike is very close to the event horizon of the BH, so the interaction between the BH spin and the DM spike should be considered \citep{1999PhRvL..83.1719G}; (3) Observational evidence from the Event Horizon Telescope (EHT) that the M87 SMBH has a spin, as well as from the LIGO gravitational wave probe that the BH has a spin, suggests that the more real BH is the Kerr BH \citep{2019ApJ...875L...1E,2016PhRvL.116f1102A}.

From deformed Schwarzschild BHs to Kerr BHs, Newman-Janis(NJ) algorithm may be a suitable and convenient method \citep{Newman:1965tw,2014PhRvD..90f4041A,2014PhLB..730...95A}. In NJ algorithm, it is necessary to solve a set of differential equations to obtain the deformed Kerr BH metric. The detailed derivation of the metric of deformed Kerr BH is as follows. 
In the deformed Schwarzschild BH metric (\ref{DM-S16}), the space-time metric is described by the coordinates $(t,r,\theta,\phi)$, the time coordinate $t$ and space coordinates $(r,\theta,\phi)$ correspond to the Minkowski situation in flat space-time. In NJ algorithm, we need advanced null coordinate $(u,r,\theta,\phi)$ instead of Schwarzschild coordinate $(t,r,\theta,\phi)$. The connection is as follows
\begin{equation}
du=dt-\dfrac{1}{F(r)G(r)}dr.
\label{NJA1}
\end{equation}
In the null tetrad, there are four basis vectors, and they are $l^{\mu}$, $n^{\mu}$, $m^{\mu}$ and $\overline{m}^{\mu}$. These basis vectors satisfy two conditions: (1) the magnitude of the basis vector is $1$; (2) they satisfy the orthogonal condition.
In the null tetrad, the space-time metric can be represented by linear combinations of basis vectors, and the inverse metric is 
\begin{equation}
g^{\mu\nu}=-l^{\mu}n^{\nu}-l^{\nu}n^{\mu}+m^{\mu}\overline{m}^{\nu}+m^{\nu}\overline{m}^{\mu}.
\label{NJA2}
\end{equation}
For the metric of a deformed BH in DM spike, the basis vectors are 
\begin{equation}
l^{\mu}=\delta^{\mu}_{r},$$$$
n^{\mu}=\sqrt{\dfrac{F(r)}{G(r)}}\delta^{\mu}_{\mu}-\dfrac{F(r)}{2}\delta^{\mu}_{r},$$$$
m^{\mu}=\dfrac{1}{\sqrt{2}r}\delta^{\mu}_{\theta}+\dfrac{i}{\sqrt{2}r \sin\theta}\delta^{\mu}_{\phi},$$$$
\overline{m}^{\mu}=\dfrac{1}{\sqrt{2}r}\delta^{\mu}_{\theta}-\dfrac{i}{\sqrt{2}r \sin\theta}\delta^{\mu}_{\phi}.
\label{NJA3}
\end{equation}
Under the null trade, the space-time coordinates between different observers satisfy the complex transformation, which is the main means for us to derive the metric of the deformed kerr BH, and these transformations are $u\rightarrow u-ia\cos\theta$, $r\rightarrow r+ia\cos\theta$. It turns out that this transformation is equivalent to a rotation in Schwarzschild coordinates. When the complex transformation is considered, the metric coefficient of the deformed BH becomes more complicated, and the metric coefficient becomes a function of $(r,\theta,a)$. 
After the complex transformation, the metric coefficients change as follows: $f(r)\rightarrow \widetilde{F}(r,\theta,a)$, $g(r)\rightarrow \widetilde{G}(r,\theta,a)$ and $h(r)(=r^{2})\rightarrow \widetilde{\Psi}(r,\theta,a)$.
The basis vectors are
\begin{equation}
l^{\mu}=\delta^{\mu}_{r},$$$$
n^{\mu}=\sqrt{\dfrac{\widetilde{G}}{\widetilde{F}}}\delta^{\mu}_{\mu}-\dfrac{\widetilde{F}}{2}\delta^{\mu}_{r},$$$$
m^{\mu}=\dfrac{1}{\sqrt{2\widetilde{\Psi}}}(\delta^{\mu}_{\theta}+ia \sin\theta(\delta^{\mu}_{\mu}-\delta^{\mu}_{r})+\dfrac{i}{\sin\theta}\delta^{\mu}_{\phi}),$$$$
\overline{m}^{\mu}=\dfrac{1}{\sqrt{2\widetilde{\Psi}}}(\delta^{\mu}_{\theta}-ia \sin\theta(\delta^{\mu}_{\mu}-\delta^{\mu}_{r})-\dfrac{i}{\sin\theta}\delta^{\mu}_{\phi}).
\label{NJA5}
\end{equation}
Therefore, based on the metric expression of (\ref{NJA2}), we obtain the contravariant metric 
\begin{equation}
g^{uu}=\dfrac{a^{2}\sin^{2}\theta}{\widetilde{\Psi}},~~~~~~g^{\theta\theta}=\dfrac{1}{\widetilde{\Psi}},$$$$
g^{ur}=g^{ru}=\sqrt{\dfrac{\widetilde{G}}{\widetilde{F}}}-\dfrac{a^{2}\sin^{2}\theta}{\widetilde{\Psi}},$$$$
g^{\phi\phi}=\dfrac{1}{\widetilde{\Psi}\sin^{2}\theta},~~~~~~
g^{u\phi}=g^{\phi u}=\dfrac{a}{\widetilde{\Psi}},$$$$
g^{r\phi}=g^{\phi r}=\dfrac{a}{\widetilde{\Psi}}, ~~~~~~g^{rr}=\widetilde{G}+\dfrac{a^{2}\sin^{2}\theta}{\widetilde{\Psi}}.
\label{NJA6}
\end{equation}
Based on these non-zero contravariant metric, the deformed Kerr BH in Eddington-Finkelstein coordinates (EFC) is given by
\begin{widetext}
\begin{equation}
ds^{2}=-\widetilde{F}du^{2}+2\sqrt{\dfrac{\widetilde{F}}{\widetilde{G}}}dudr+2a\sin^{2}\theta \big(\sqrt{\dfrac{\widetilde{F}}{\widetilde{G}}}+\widetilde{F} \big)dud\phi-2a\sin^{2}\theta\sqrt{\dfrac{\widetilde{F}}{\widetilde{G}}}drd\phi+\widetilde{\Psi} d\theta^{2}
-\sin^{2}\theta \big[-\widetilde{\Psi}+a^{2}\sin^{2}\theta \big(2\sqrt{\dfrac{\widetilde{F}}{\widetilde{G}}}+\widetilde{F}\big) \big]d\phi^{2}.
\label{NJA7}
\end{equation}
\end{widetext}
From BH physics, the deformed Kerr BH metric can be transformed from EFC to Boyer-Lindquist coordinates (BLC$'$s) by mathematical transformation. These transformations are
\begin{equation}
du=dt-\dfrac{r^{2}\sqrt{\dfrac{F}{G}}+a^{2}}{r^{2}F+a^{2}}dr,$$$$
d\phi=d\phi-\dfrac{a}{r^{2}F+a^{2}}dr.
\label{NJA8}
\end{equation}
Through proper analysis, it is found that the relationship between $\widetilde{F}(r,\theta,a)$, $\widetilde{G}(r,\theta,a)$ and $\widetilde{\Psi}(r,\theta,a)$ are
\begin{equation}
\widetilde{F}(r,\theta,a)=-\dfrac{r^{2}F+a^{2}\cos^{2}\theta}{r^{2}\sqrt{\dfrac{F}{G}}+a^{2}\cos^{2}\theta}\widetilde{\Psi}(r,\theta,a),$$$$
\widetilde{G}(r,\theta,a)=-\dfrac{r^{2}F+a^{2}\cos^{2}\theta}{\widetilde{\Psi}(r,\theta,a)}.
\label{NJA9}
\end{equation}
To make the deformed Kerr BH metric more compact, we can make some notation stipulations: $\Sigma^{2}=K+a^{2}\cos^{2}\theta$, $K=r^{2}\sqrt{\dfrac{F}{G}}$, $\Delta=r^{2}F+a^{2}$ and $A=(K+a^{2})^{2}-a^{2}\Delta \sin^{2}\theta$. The deformed Kerr BH metric is
\begin{widetext}
\begin{equation}
ds^{2}=-\dfrac{\widetilde{\Psi}}{\Sigma^{2}}\Big(1-\dfrac{K-r^{2}F}{\Sigma^{2}}\Big)dt^{2}+\dfrac{\widetilde{\Psi}}{\Delta}dr^{2}-\dfrac{2a(K-r^{2}F)\sin^{2}\theta \widetilde{\Psi}}{\Sigma^{4}}dtd\phi+\widetilde{\Psi}d\theta^{2}+\dfrac{\widetilde{\Psi} A \sin^{2}\theta}{\Sigma^{4}}d\phi^{2}.
\label{NJA10}
\end{equation}
\end{widetext}

The metric (\ref{NJA10}) of a deformed BH is a universal form, but it$'$s not necessarily a BH solution, because we don't yet know if it satisfies Einstein$'$s field equation. Therefore, the  deformed BH metric (\ref{NJA10}) should satisfy two conditions: $G_{r\theta}=0$ and $G_{\mu\nu}=8\pi T_{\mu\nu}$, and obtain the following equations($y=\cos\theta$)
\begin{widetext}
\begin{equation}
(K+a^{2}y^{2})^{2}(3\widetilde{\Psi}_{,r}\widetilde{\Psi}_{,y^{2}}-2\widetilde{\Psi}\widetilde{\Psi}_{,ry^{2}})=3a^{2}K_{,r}\widetilde{\Psi}^{2},
\label{NJA11}
\end{equation}
and 
\begin{equation}
\widetilde{\Psi}[K^{2}_{,r}+K(2-K_{,rr})-a^{2}y^{2}(2+K_{,rr})]+(K+a^{2}y^{2})(4y^{2}\widetilde{\Psi}_{,y^{2}}-K_{,r}\Psi_{,r})=0.
\label{NJA12}
\end{equation}
\end{widetext}
Equations (\ref{NJA11}) and (\ref{NJA12}) are complex partial differential equations, and it is difficult to obtain the exact solution in general situation ($F\neq G$). 
Fortunately, for a deformed BH (\ref{DM-S16}), the metric satisfies the condition $F=G$. 
If the Kerr BH metric is taken as the boundary condition, the solutions of Equations (\ref{NJA11}) and (\ref{NJA12}) can be obtained as
\begin{equation}
K=r^{2}, $$$$
\widetilde{\Psi}=r^{2}+a^{2}\cos^{2}\theta.
\label{NJA13}
\end{equation}
Then, the final deformed Kerr BH metric in DM spike can be obtained 
\begin{widetext}
\begin{equation}
\begin{aligned}
&ds^{2}=-\dfrac{r^{2}G+a^{2}\cos^{2}\theta}{\Sigma^{2}}dt^{2}+\dfrac{\Sigma^{2}}{\Delta}dr^{2}-\dfrac{2ar^{2}\sin^{2}\theta(1-G)}{\Sigma^{2}}d\phi dt\\
&+\Sigma^{2}d\theta^{2}+\Sigma^{2}\sin^{2}\theta [1+a^{2}\sin^{2}\theta\dfrac{2r^{2}-r^{2}G+a^{2}\cos^{2}\theta}{\Sigma^{2}}d\phi^{2}],
\label{K1}
\end{aligned}
\end{equation}
\end{widetext}
where $\Sigma^{2}=r^{2}+a^{2}\cos^{2}\theta$ and $\Delta=r^{2}G+a^{2}$. The specific analytic form of $\Delta$ is
\begin{widetext}
\begin{equation}
\Delta=r^{2}G+a^{2}=r^{\dfrac{48\pi R^{\alpha}_{\rm sp}\rho_{\rm R}(kR_{S})^{3-\alpha}}{\alpha(\alpha-1)(\alpha-2)(\alpha-3)}+2}\times {\rm exp}\Big(-\dfrac{8\pi R^{\alpha}_{\rm sp}\rho_{\rm R}(kR_{S})^{3}}{\alpha^{2}}r^{-\alpha}+$$$$
\dfrac{24\pi R^{\alpha}_{\rm sp}\rho_{\rm R}(kR_{S})^{2}}{(\alpha-1)^{2}}r^{1-\alpha}-\dfrac{24\pi R^{\alpha}_{\rm sp}\rho_{\rm R}kR_{S}}{(\alpha-2)^{2}}r^{2-\alpha}+\dfrac{8\pi R^{\alpha}_{\rm sp}\rho_{\rm R}}{(\alpha-3)^{2}}r^{3-\alpha}\Big)-2r M_{\rm BH}+a^{2}.
\label{metric-K3}
\end{equation}
\end{widetext}
The deformed Kerr BH metric (\ref{K1}) describe the DM spike halo influence on center BH space-time. It tells us how DM interacts with BHs. The space-time metric (\ref{K1}) can be classified as follows:

$(1)$ If the DM spike is absent $\rho_{\rm R}=0$, this deformed space-time metric reduces to Kerr BH. In the real universe, Kerr BHs do not exist, because any BH must have matter nearby. But a Kerr BH is a pretty good approximation. At the same time, Kerr BH is also a widely studied BH solution. 

$(2)$ If the BH spin $a=0$ for deformed Kerr BH metric, the (\ref{K1}) reduces to (\ref{DM-S16}), which describe the DM spike halo influence on the Schwarzschild BH. 

$(3)$ If the DM spike halo turns into a DM halo, that is $R_{S}=0$, $\alpha=\alpha_{ini}$, $R_{\rm sp}=R_{0}$ and $\rho_{\rm R}=\rho_{\rm 0}$ in line element (\ref{K1}). For NFW halo situation, the $\alpha_{ini}=1$, and Uniform Density (SIDM and FDM) situation corresponding to $\alpha_{ini}=0$. This type of deformed BH metric describes the space-time metric away from the BH, where the DM halo dominates.

$(4)$ When parameter $k=4$, the deformed BH metric (\ref{K1}) represents the influence of the GS DM spike on the Kerr BH, and the deformed BH metric (\ref{K1}) is applicable to describe the interaction between BH and DM at a large distance. When $k=2$, the deformed metric (\ref{K1}) represents the influence of the SFW DM spike on the Kerr BH, and the BH metric (\ref{K1}) is applicable to describe the interaction between BH and DM near the event horizon.

$(5)$ For the NFW profile corresponding to the CDM model, the corresponding power index of DM spike is $\alpha=7/3$, and then the deformed Kerr BH metric (\ref{K1}) describes the influence of CDM particles on Kerr BH space-time. For the distribution corresponding to the SIDM model (or the TF profile corresponding to the FDM model), the corresponding DM spike power index is $\alpha=3/2$ (Uniform Density), and then the deformed Kerr BH metric (\ref{K1}) describes the influence of SIDM particles and ultralilight particles on the Kerr BH.

\section{BH properties}
\label{pro-}
In the section \ref{metrics}, we obtained the deformed BH metric that describes the effect of the DM spike on the Schwarzschild BH. In section \ref{metric-kerr}, we obtained the deformed BH metric which describes the effect of DM spike on Kerr BH. Of course, these deformed BH metrics are approximate, and they capture the main features of the interaction between DM and BHs, while ignoring the microscopic nature of the interaction. According to deformed Kerr BH metric (\ref{K1}), we will detailed analyzing the distribution of DM to change the properties of BHs, such as DM spike how to change the size of the BH horizon, how DM spike changes geometric properties of BH ergospheres, DM spike how to change the stationary limit surface properties, DM how to change the BH singularity, etc.

\begin{figure}[htbp]
  \centering
   \includegraphics[scale=0.35]{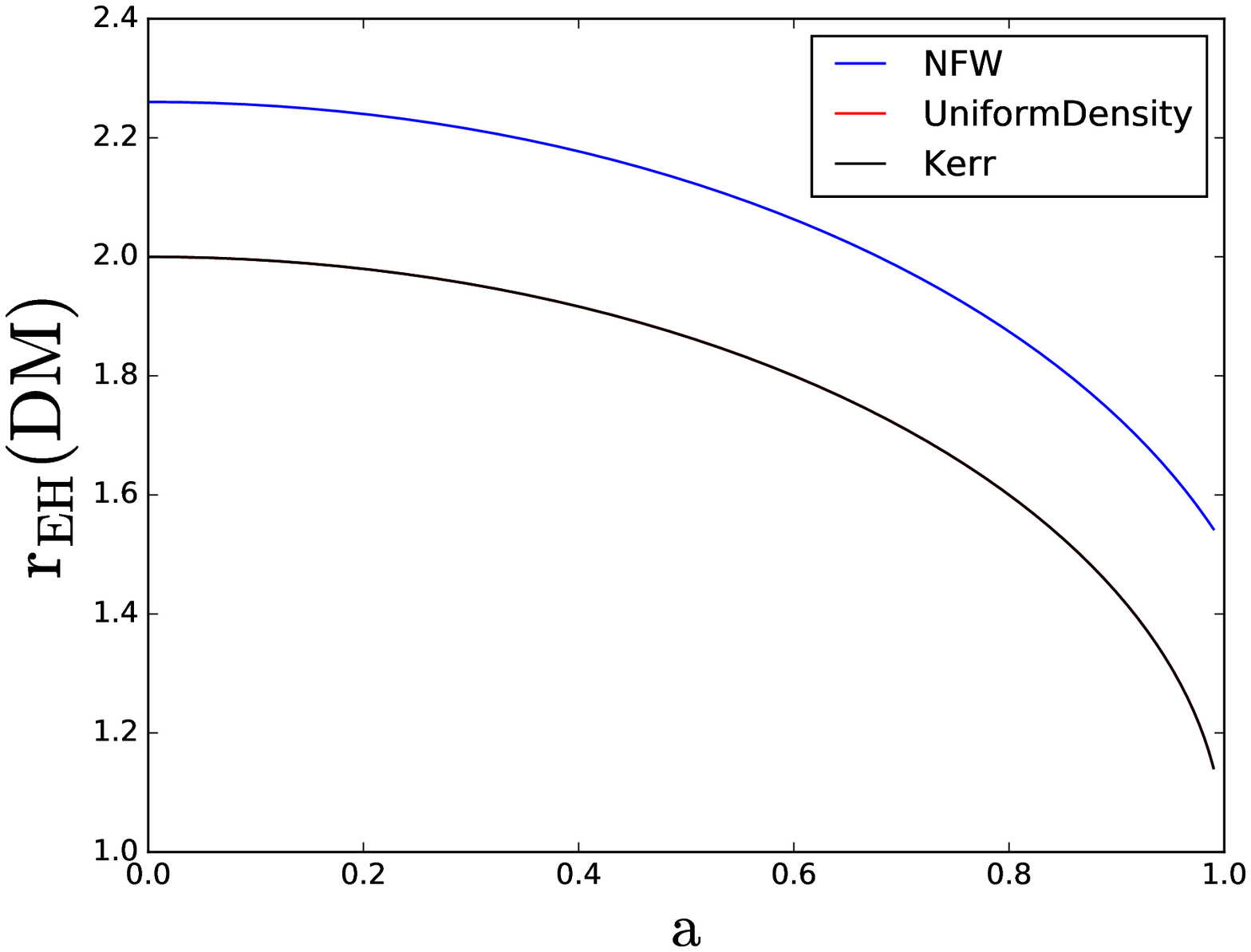}
   \includegraphics[scale=0.35]{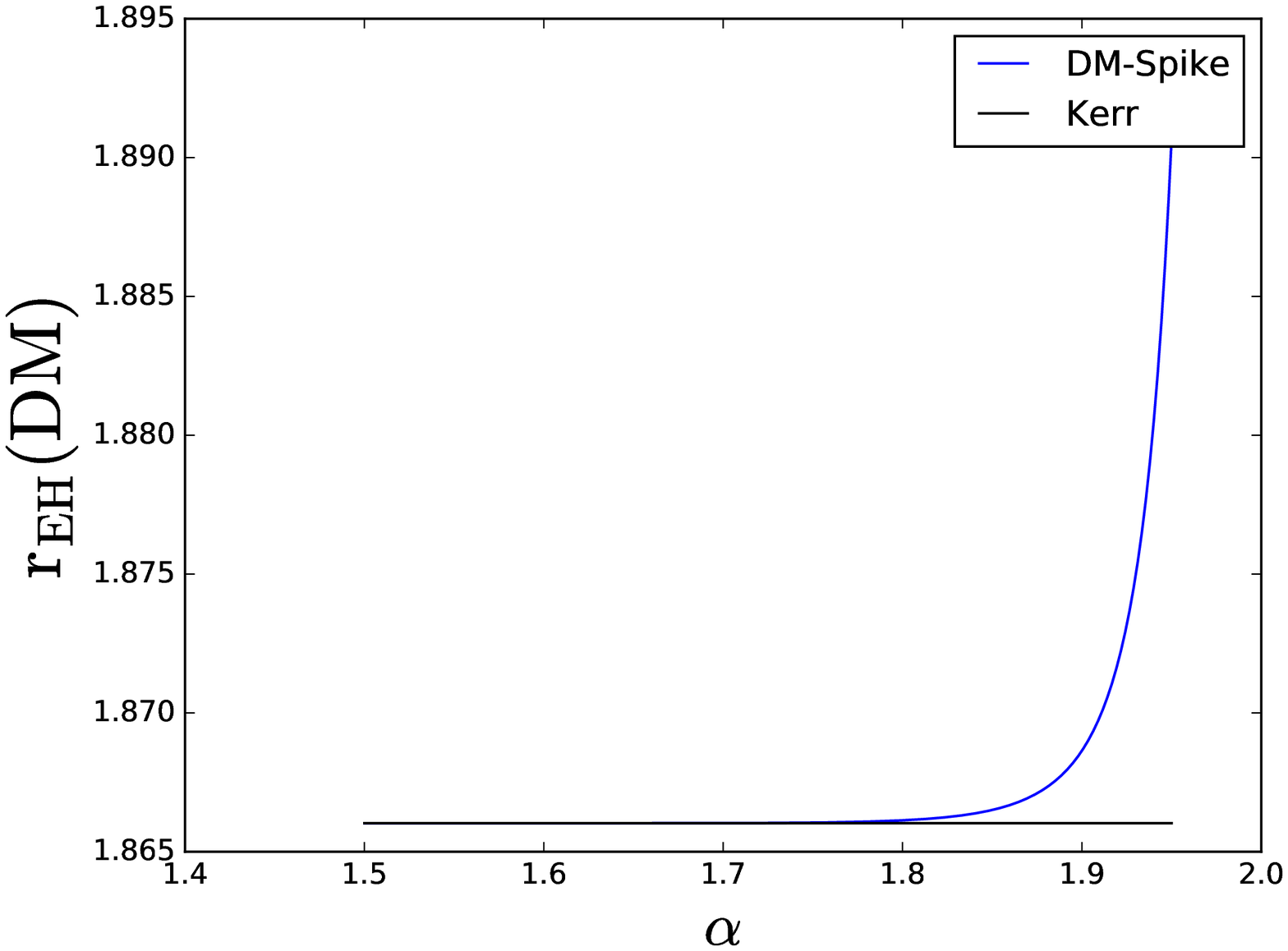}
   \includegraphics[scale=0.35]{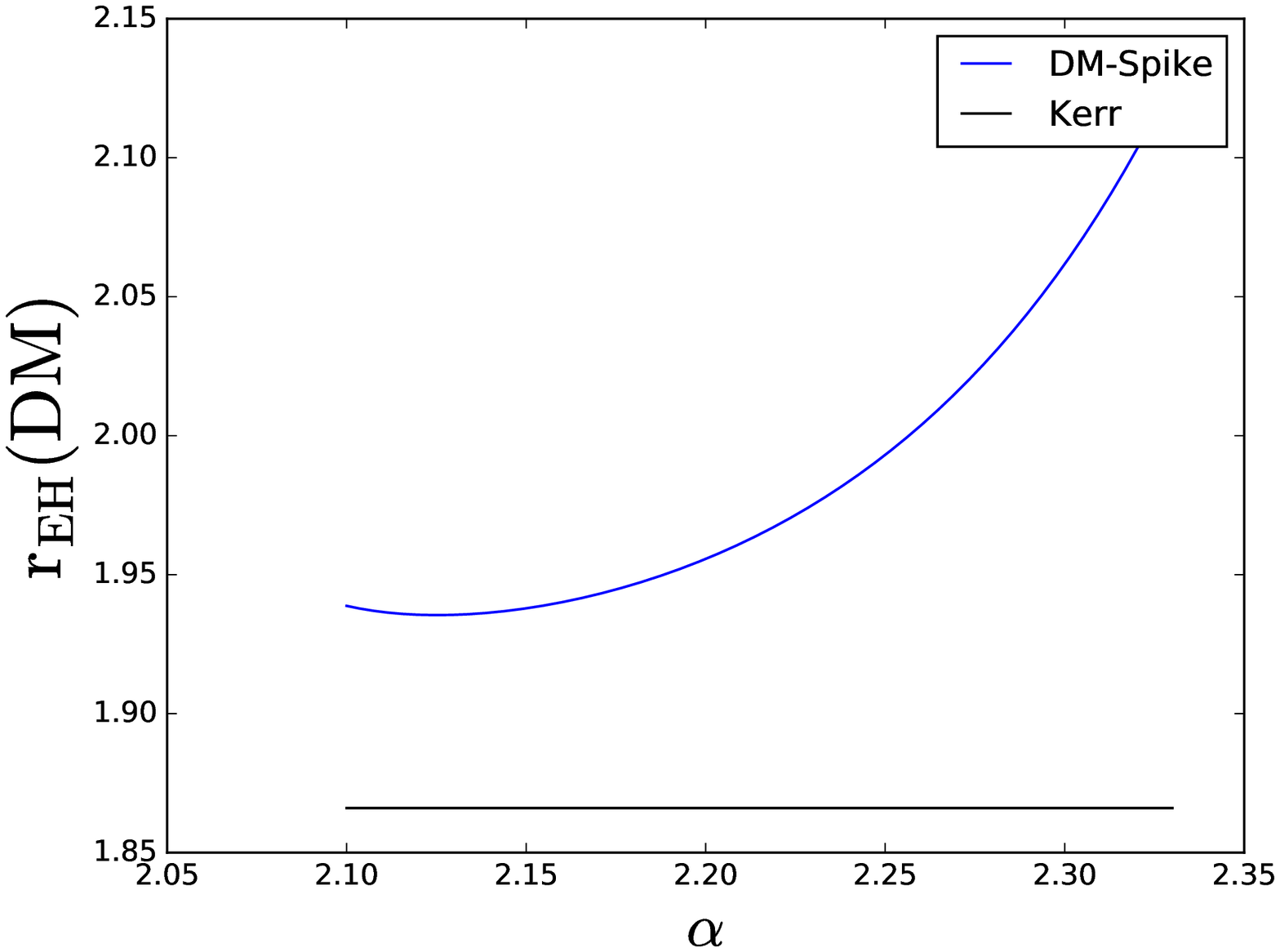}
   \includegraphics[scale=0.35]{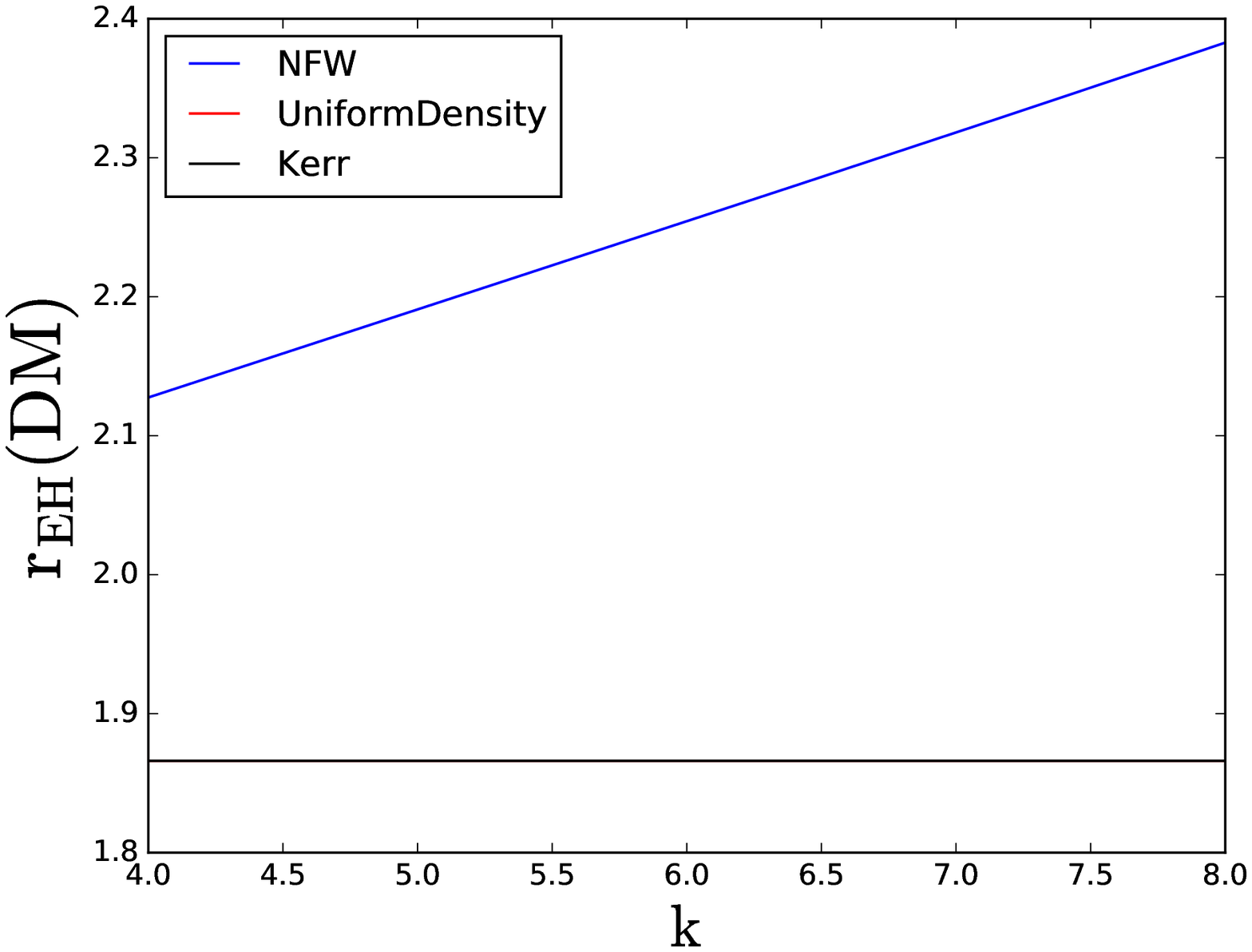}
   \caption{Deformed BH event horizon changes with BH spin $a$, DM spike power-law index $\alpha$ and post-newtonian approximation parameter $k$.}
  \label{EH100}
\end{figure}

\textbf{Event horizon}. 
The event horizon (EH) is the surface of a BH. Inside and outside the BH$'$s event horizon, space-time behaves very differently. Generally speaking, Schwarzschild BHs have only one event horizon, which is uniquely determined by the mass of the BH. The Kerr BH has two event horizons, the event horizon and the Cauchy horizon, which are determined by the mass and spin of the BH. So what happens to the deformed BH metric (\ref{K1})? According to previous results, it is generally believed that the DM spike does not change the number of deformed BH event horizons, it just changes the size of these event horizons. Then, we analyze the space-time metric of the deformed Kerr BH metric  (\ref{K1}) in order to learn more about the changes in the event horizon of the BH. We mainly consider the DM spike corresponding to the CDM model and the SIDM model, and the main results are shown in Fig.\ref{EH100}. We find the following results: (1) compared with the DM halo, the DM spike increases the size of the BH$'$s event horizon more obviously, and such enhancement effect is likely to be detected; (2) the spike distribution of DM has different effects on the radius of the BH event horizon. The CDM model is the most significant, while the SIDM model is much weaker; (3) when the DM spike structure is not considered, the results here are basically consistent with those in the literature. 

BH event horizon size depends on DM spike parameters has special physical significance. In BH physics, it is generally believed that after a massive star forms a BH, only the information such as mass, angular momentum and charge is left, and the rest information such as the structure of matter will all disappear, which is the famous no hair theorem. However, this is not the case with deformed BH (\ref{K1}). As the previous analysis suggests, the size of the BH$'$s event horizon carries information about the distribution of DM, making it the source of the BH$'$s space-time. Unfortunately, the deformed Kerr BH we get here is only an approximate solution. If it is an exact solution, this result will have important physical significance.

\begin{figure}[htbp]
  \centering
   \includegraphics[scale=0.35]{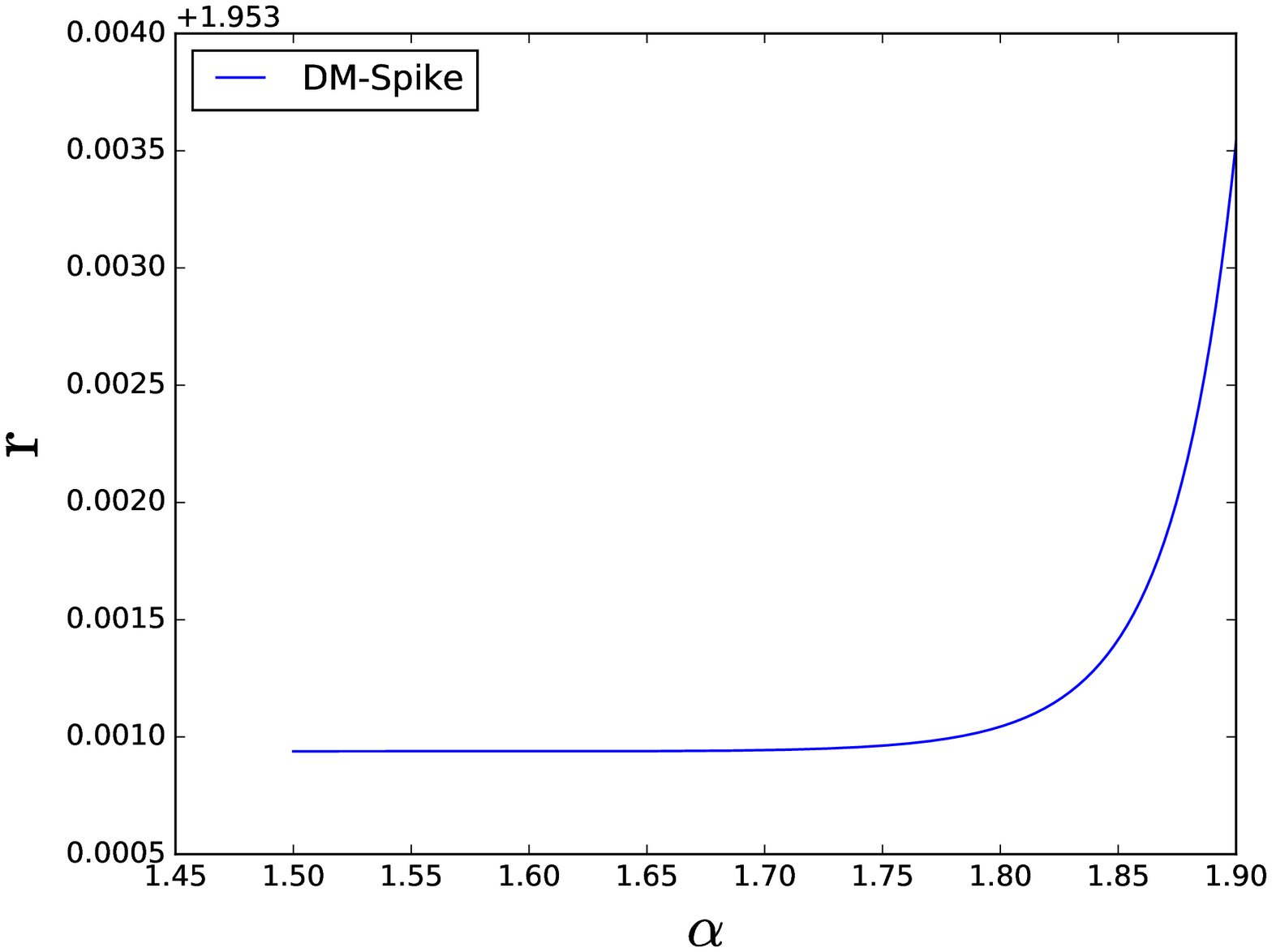}
   \includegraphics[scale=0.35]{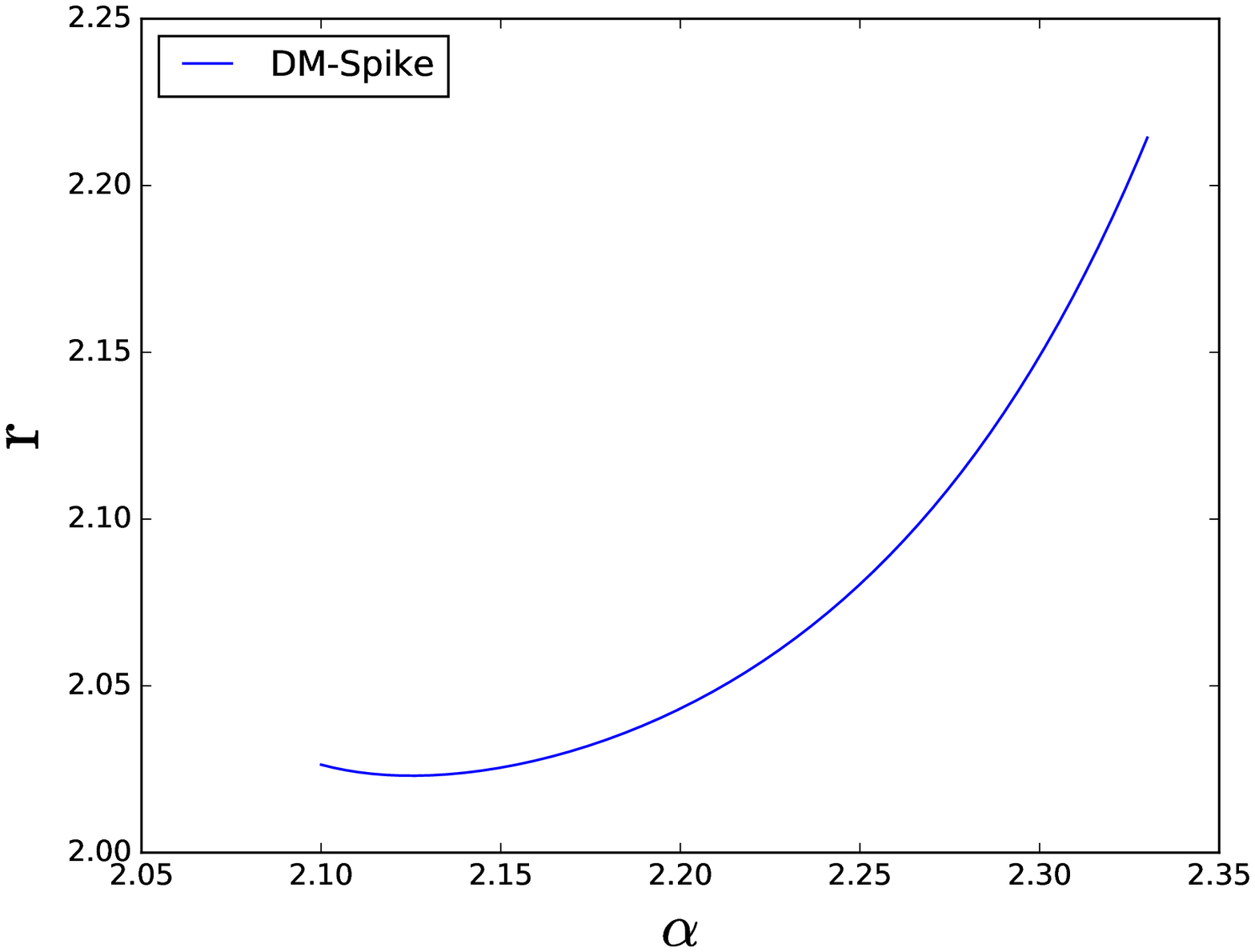}
   \includegraphics[scale=0.40]{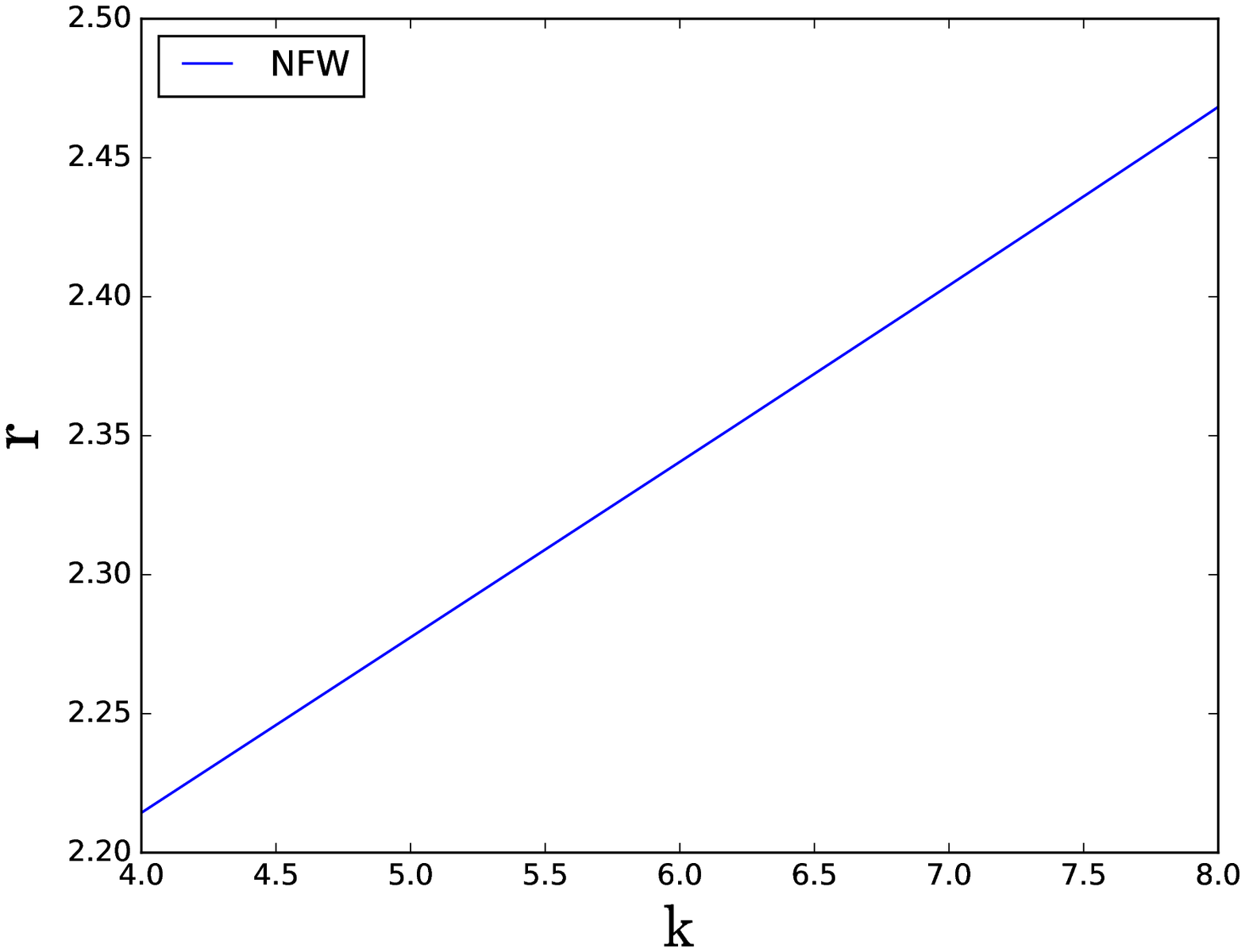}
   \caption{Deformed BH SIS changes with DM spike power-law index $\alpha$ and post-newtonian approximation parameter $k$.}
  \label{EH200}
\end{figure}

\textbf{Stationary limit surfaces and ergosphere}. 
Stationary limit surfaces (SIS) are a unique property of a spinning BH. They are curved surfaces from which photons cannot reach the observer when they are emitted. When the spin of the BH is zero, the stationary limit surfaces coincides with the event horizon of the BH. When the spin of the BH is non-zero, two static limit planes will appear, and these static limit planes will be separated from the event horizon of the BH. The situation is similar for the deformed BH metric (\ref{K1}). The SIS is determined by condition $g_{tt}=r^{2}G+a^{2}\cos^{2}\theta=0$. 
By numerically calculating this equation and combining with the results of the event horizon of the deformed BH, we mainly get the following results (Fig.\ref{EH200}): (1) there are two SIS of the deformed BH, namely $r^{-}_{SIS}$ and $r^{+}_{SIS}$, which change with the BH mass, BH spin and DM spike parameters; (2) the spike distribution of CDM enhanced SIS most significantly, the spike distribution of SIDM enhanced SIS the next, and the pure DM halo was the weakest; (3) between the event horizon and the $r^{+}_{SIS}$ exist regions of the ergeosphere, like the Kerr BH, in which there are negative energy orbits, making it possible to extract the BH$'$s energy; (4) the size and structure of the ergeosphere are affected by the spike structure of DM.

\textbf{Singularity structure}. 
According to the BH singularity theorem, when a massive star through gravitational collapse to form a BH, there will be a singularity in the center of the BH. The singularity of Schwarzschild BHs is represented by a point, and the singularity of Kerr BHs is represented by a ring. In previous work, it was found that the distribution of dark matter changes the specific behaviour of the BH singularity. It can be speculated that in the  deformed BH metric (\ref{K1}), the singularity of BH will also change accordingly. The singularity of BH can be analyzed by calculating the Kretsmann scalar $R$, and its expression can be obtained by calculation
\begin{equation}
R=R^{\mu\nu\beta\delta}R_{\mu\nu\beta\delta}=\dfrac{Z(g(r),a,M_{\rm BH},\theta)}{\Sigma^{12}},
\label{PRO-S10}
\end{equation}
where $Z(g(r),a,M_{\rm BH},\theta)$ are the polynomial functions of metric coefficient $g(r)$, BH spin $a$ and BH mass $M_{\rm BH}$. It is found that the singularity of the BH is the same as Kerr BH, the position and size of the singularity ring do not change. This may be due to the equivalence of $f(r)=g(r)$ in metric (\ref{K1}) of the deformed BH. Our main conclusions are as follows: (1) 
in the deformed BHs metric (\ref{K1}), the singularity of BH is the same as Kerr BH, which is an singularity ring of radius $a$; (2) the Kretsmann scalar of the deformed BH is determined by the BH mass, BH spin and DM spike; (3) the Kretsmann scalar near the BH$'$s singularity ring is related to the distribution of matter throughout the universe, although the influence of distant objects is very small; (4) when the DM spike disappears $\rho_{\rm R}=0$, (\ref{PRO-S10}) degenerates to the Kerr BH case.

\section{Summary}
\label{sum-}
In this work, we obtain the deformed BH metric in DM spike, and analysis the properties of these BH. We found that the DM spike can largely enhance the center black hole EH and SIS. Our main results are as follows: (1) Under the spike distribution of DM, we get the metric of a deformed Schwarzschild BH (\ref{DM-S16}) and deformed Kerr BH (\ref{K1}). In this deformed BH metric, different exponents $\alpha$ correspond to different DM models. Among them, we are most interested in the CDM model ($\alpha=7/3$) and SIDM model ($\alpha=3/2$). We only need to put the DM parameters into (\ref{DM-S16}) and (\ref{K1}) respectively to get the corresponding metric of the deformed BH.
(2) we calculate the basic properties of the deformed BH metric (\ref{K1}), such as the BH event horizon, SIS, and BH singularity. We found that, compared with the DM halo, the existence of DM spike greatly improves the change of BH metric. These results contribute to our understanding of how DM spikes interact with BHs.

\begin{acknowledgments}
We acknowledge the anonymous referee for a constructive report that has significantly improved this paper. We acknowledge the financial support from the China Postdoctoral Science Foundation funded project under grants No. 2019M650846.
\end{acknowledgments}

\nocite{*}

\bibliography{apssamp}% Produces the bibliography via BibTeX.

\end{document}